\documentstyle{article}

\newlength{\defaultparindent}
\newenvironment{Default Paragraph Font}{}{}
\input{tcilatex}

\begin{document}

A Mode of Unifying Gravitation and Strong Interactions Proposed by the Model
of Expansive Nondecelerative Universe

Miroslav Sukenik and Jozef Sima

Slovak Technical University, Radlinskeho 9, 812 37 Bratislava, Slovakia

Abstract

The model of Expansive Nondecelerative Universe exploiting the Vaidya
metrics is used as a tool for unification of gravitation and strong
interactions. The proposed approach stems from the capability to localize
the energy of gravitational field and enables to reach a certain level in
unifying the general theory of relativity and quantum chromodynamics. A
relationship between the energy binding quarks and gravitational energy of
virtual black holes is rationalized.

In the model of Expansive Decelerative Universe (ENU) continuous matter and
gravitational field creations occur simultaneously. Due to the negative
value of the latter, the total energy of the Universe is exactly of zero
value [1]. Such a Universe can expand without limitation by the velocity of
light [2, 3]. Based on the total zero energy value it can be postulated that
also the total charge, linear momentum and angular momentum of the Universe
are of zero value. This is why the gravitation is understood only as one of
the forms of the matter creation which can manifest itself in cases when the
density of gravitational energy of any body exceeds the critical
gravitational energy density. Further, we suppose that energies of all the
fundamental physical interactions must be compensated by the negative energy
of gravitational field. This is a key point allowing to propose a mode of
unification of the fundamental physical interactions.

In our previous paper [3] it was evidenced that within the first
approximation the Tolman's relation in ENU reads as follows

$\epsilon _{g}=-\frac{R.c^{4}}{8\pi .G}=-\frac{3m.c^{2}}{4\pi .a.r^{2}}$ \ \
\ \ \ \ \ \ \ \ \ \ \ \ \ \ \ \ \ \ \ \ \ \ \ \ \ (1)

where $\varepsilon _{g}$ is the density of the gravitational energy emitted
by a body with the mass $m$ at the distance $r,R$ denotes the scalar
curvature (contrary to a more frequently used Schwarzschild metrics, in the
Vaidya metrics $R\neq $ 0 [4] also outside the body).

For the energy binding quarks it holds that the higher distance between
them, the higher energy of their binding (the energy approaches the zero
value in a limiting case of zero distance). This fact can be expressed by
relation

$E_{b}=\frac{\hbar .c}{a}.\frac{r^{2}}{l_{Pc}^{2}}$ \ \ \ \ \ \ \ \ \ \ \ \
\ \ \ \ \ \ \ \ \ \ \ \ \ \ \ \ \ \ \ \ \ \ \ \ \ \ \ \ (2)

in which $E_{b}$ is the binding energy of a quarks couple, $r$ is their
distance, $\hbar .c/a$ represents the minimum possible energy, i.e. the
energy of a photon of the wavelength $a$, $l_{Pc}$ is the Planck distance
defined as

$l_{Pc}=\left( \frac{G.\hbar }{c^{3}}\right) ^{1/2}\approx 10^{-35}m$ \ \ \
\ \ \ \ \ \ \ \ \ \ \ \ \ \ \ \ \ \ \ \ (3)

and $a$ is the gauge factor that at present

$a\approx 10^{26}m$ \ \ \ \ \ \ \ \ \ \ \ \ \ \ \ \ \ \ \ \ \ \ \ \ \ \ \ \
\ \ \ \ \ \ \ \ \ \ \ \ \ \ \ (4)

In some theoretical concepts a linear relationship between the binding
energy and the distance is introduced, it should be pointed out that the
validity of relation (2) postulated by us in [5] was theoretically proved in
an independent way [6].

As follows from (2) there are two limiting values of binding energy. In
cases when $r$ = $l_{Pc}$, the binding energy is of the minimum value and
contrary, when $r=a$, this energy is equal to $M_{u}$.$c^{2}$[5] which
represents the maximum possible value ($M_{u}$ is the mass of the Universe).
In actual cases $r$ $\simeq $10$^{-15}$m (the range of nuclear forces), the
binding energy of quarks approximates $E_{b}$ $\simeq $10$^{-11}$ J. This
value is in excellent agreement with the kinetic energy of $\pi ^{+}$ mesons
(200 MeV) determined by their scattering on protons accompanied by
subsequent resonance creation.

When attempting to unify the general theory of relativity and quantum
chromodynamics one must realize the disproportion between a comparatively
small mass of the quarks and substantial energy binding them together. Based
on the generally accepted concept of virtual black holes we supposed that
the energy binding two quarks being in the distance $r$ is equal to the
energy of the gravitational field of a virtual black hole of diameter $r$.
The largest is the black hole, the largest is its gravitational energy
which, in turn, is identical to the binding energy between two quarks. This
phenomenon can be described by the term asymptotic freedom.

For the binding energy it must then hold

$E_{b}=\left| E_{g}\right| =\left| \int \epsilon _{g}dV\right| $ \ \ \ \ \ \
\ \ \ \ \ \ \ \ \ \ \ \ \ \ \ \ \ \ \ \ \ \ (5)

where $\epsilon _{g}$ is the energy density of a virtual black hole and $%
E_{g}$ is the energy of its gravitational field. It follows from (1) and (5)
that

$E_{b}=\frac{m_{bh}.c^{2}.r}{a}$ \ \ \ \ \ \ \ \ \ \ \ \ \ \ \ \ \ \ \ \ \ \
\ \ \ \ \ \ \ \ \ \ \ \ \ \ \ \ \ \ (6)

where $m_{bh}$ is the black hole mass, $r$ is its gravitational diameter
representing at the same time the distance between two quarks. The mass of
black hole can be calculated as

$m_{bh}=\frac{r.c^{2}}{2G}$ \ \ \ \ \ \ \ \ \ \ \ \ \ \ \ \ \ \ \ \ \ \ \ \
\ \ \ \ \ \ \ \ \ \ \ \ \ \ \ \ \ \ \ (7)

and substitution (6) into (7) leads finally to

$E_{b}=\frac{c^{4}.r^{2}}{2G.a}$ \ \ \ \ \ \ \ \ \ \ \ \ \ \ \ \ \ \ \ \ \ \
\ \ \ \ \ \ \ \ \ \ \ \ \ \ \ \ \ \ \ \ \ (8)

Multiplying both the denominator and numerator by $\hbar $ we get relation
(2) evidencing the unification of the general theory of relativity and
quantum chromodynamics.

\bigskip For the effective gravitational range $r_{ef(g)}$ of a particle
with the mass $m$ it holds [8]

$r_{ef(g)}=\left( R_{g(m)}.a\right) ^{1/2}$ \ \ \ \ \ \ \ \ \ \ \ \ \ \ \ \
\ \ \ \ \ \ \ \ \ \ \ \ \ \ \ (9)

where $R_{g(m)} $  is the gravitational diameter of the particle. It can be
subsequently shown that the absolute value of the gravitational energy of a
virtual black hole with the radius $r_{ef(g)} $  is equal exactly to the
rest energy of the particle, i.e. to $m.c^{2} .$

Conclusions

The present contribution is a tentative proposal of a new direction that
might lead to unification of the fundamental physical interactions. As a
starting point, the ENU model exploiting the Vaidya metrics and enabling to
localize gravitational energy is used. The paper provides conclusions on
relation between strong interactions and gravitation as the first
approximation to the elaborated issue, the next steps are expected to
involve time-dependent modified Vaidya, Reissner-Nordstrm and Kerr-Newman
metrics.

\bigskip

References

1.Hawking, S.: A Brief History of Time: From the Big Bang to Black Holes,
Bantam Books, New York, p. 129

2.Skalsky, M.Sukenik: Astrophys. Space Sci.,178 (1991) 169

3.J. Sima, M. Sukenik: General Relativity and Quantum Cosmology, Preprint
in: US National Science Foundation, E-print archive: gr-qc@xxx.land.gov.,
paper 9903090 (1999)

4.P.C. Vaidya: Proc. Indian Acad. Sci., A33 (1951) 264

5.Skalsky, M.Sukenik: Astrophys. Space Sci.,236 (1996) 295

6.Shipov, G.I.: The Theory of Physical Vacuum, NT Center, Moscow, 1993, p.
183 (in Russian)

7.V. Ullmann: Gravity, Black Holes and the Physics of Time-Space,
Czechoslovak Astronomic Society, CSAV, Ostrava, (1986), p. 257 (in Czech)

8.J. Sima, M. Sukenik, M. Sukenikova, General Relativity and Quantum
Cosmology, Preprint in: US National Science Foundation, E-print archive:
gr-qc@xxx.land.gov., paper 9910094 (1999)

\end{document}